
\input amstex
 \documentstyle{amsppt}
 \nologo
 \magnification=\magstep1
 \topmatter
 \title
$\qquad\qquad\qquad\qquad\qquad\qquad\qquad\qquad\qquad\qquad\qquad\qquad
 \text{UNC-MATH-91/3}$\\ Homological (ghost) approach to constrained
 Hamiltonian systems\endtitle
 \author Jim Stasheff\endauthor
 \affil Department of Mathematics\\University of North
 Carolina at Chapel Hill\endaffil
 \address Chapel Hill, NC 27599-3250, jds@math.unc.edu\endaddress
 \endtopmatter
 \document

\def\c{\cite}

\def\gothic{\frak}

\def\cal{\Cal}

\def\[{\lbrack}
\def\]{\rbrack}

\def\integral{\int}

\def\cite#1{[{\bf #1}]}

\def\integral{\int}


\def\FBV{\Lambda\eta^{\alpha}\otimes P \otimes \Lambda\cal P_{\alpha}}
\def\etaa{\eta^{\alpha}}
\def\etab{\eta^{\beta}}

\def\etae{\eta^{\epsilon}}

\def\pb{\cal P_{\beta}}
\def\pg{\cal P_{\gamma}}
\def\pe{\cal P_{\epsilon}}
\def\phia{\phi_{\alpha}}
\def\phib{\phi_{\beta}}

\def\strfun{C^{\gamma}_{\alpha \beta}}
\def\[{\lbrack}
\def\]{\rbrack}

\def\integral{\int}

\def\cal{\Cal}

\def\cite#1{[{\bf #1}]}

 \def\doit#1{\hskip-4em\hbox to 4em{#1\hss}}
 
 \pageno=1
 \headline={\ifnum\pageno>1 \tenbf
Ghost approach \hss Stasheff\else\hfil\fi}

        I had originally intended to give a talk on homological
reduction of first class constrained Hamiltonian systems, as
in my joint work with Henneaux, Fisch and Teitelboim \c {FHST}.  Since
the organizers have given me the ?honor? of opening the
conference, I will attempt to set that work in a larger
context, namely that of ghost techniques in mathematical physics.
\vskip2ex

        What are `ghosts' and what are they doing in physics?
The name reflects the fact that they are new, auxiliary variables
that are NOT physical, but are added to the system for computational
reasons.  An analogy familiar to many mathematicians is that of
a resolution in homological algebra - the generators added to
construct the resolution are the analogs of ghosts.  Indeed it
is more than an analogy in some cases; I first became seriously
interested in the subject when I read a preprint of Browning and
McMullan \c {BM} in which certain `anti-ghosts' were clearly identified
as generators of the Koszul resolution of an appropriate ideal.

\vskip2ex
        But I am getting ahead of my story, both conceptually
and historically.  My intention this morning is to set the
stage for a set of techniques and results which can be grouped
under the rubric of `cohomological physics', particularly BRST
cohomology.
\vskip2ex

        It will be easier to illustrate why BRST cohomology is
of interest than it will be to say what it is.  The acronym
BRST has come to be applied in mathematical physics very
widely; at times one gets the feeling it could be applied any
time one has an operator of square zero (called in physics
`nilpotence'), but I will try to hold back the sea and restrict
the term somewhat.
\vskip2ex

        First, BRS refers to Becchi, Rouet and Stora who in
1975 \c {BRS} called attention to the ``so-called Slavnov identities
which express an invariance of the Fadeev-Popov Lagrangian''.
The T refers to Tyutin who, at about the same time \c {Ty} had a
preprint on the same subject - the symmetry revealed is that of
gauge transformations.

\vskip2ex
        In pursuit of quantization of certain problems in gauge
field theory, Fadeev and Popov had modified certain Lagrangians
by introducing new non-physical variables which they called
ghosts.  Becchi, Rouet and Stora and Tyutin
discovered a transformation $s$ which had a striking behavior on
one of the Fadeev-Popov ghosts $c$:
$$
                        sc = 1/2[c,c].
$$
Stora, among others, soon recognized the resemblance to the
Maurer-Cartan form on a Lie group. Those were the days when the
fibre bundle setting for gauge field theories was just becoming
accepted (though implicit in the work of Dirac in 1931 (!) \c {D2}, full
recognition occurred around 1975 in the interaction of Yang
and Simons \c {Y}).
\vskip2ex
        The setting is this: we have a principal bundle
$$
  \aligned &P \\
 &\downarrow \\
  &M \endaligned$$

Here we invoke the now common dictionary between physics and mathematics:
$$\align
 \text{vector potential}\ & = \text{connection }A\\
\text{field strength } &= \text{curvature }F = dA + 1/2[A,A]\\
 \text{matter field  } &= \text{section of an associated vector bundle.}
\endalign$$
{}From a Lagrangian field theory point of view, the connection
$A$ is treated as a variable, so the action $S$ is a function
on $\cal A$, the space of connections on $P$.  The Yang-Mills
functional, for example,
$$
        YM(A) = \integral_{M^n}\vert F\vert^2 dvol:\cal A\rightarrow\Bbb R
$$
is in fact constant under changes in $A$, known as gauge equivalences.
It is hard to recall now, but in those days it was not so clear
what was ``the group of gauge transformations''.

\vskip2ex
          By 1982, the gauge transformation group had been
identified clearly as the group $\cal G$ of vertical automorphisms
of $P$ as a principal bundle. In a seminal paper, Bonora and
Cotta-Ramusino \c {BCR} identified the BRS transformation with the
standard Cartan-Chevelley-Eilenberg coboundary for the Lie
algebra cohomology of the gauge algebra $Lie \ \cal G$ with appropriate
coefficients. In that setting, the Fadeev-Popov ghosts can be
identified with elements of a weak dual of the Lie algebra $Lie\ \cal G$
of $\cal G$, namely the space of sections of the bundle
with fibre $\gothic g = Lie\ G$ associated to $P$ via the adjoint
action. (If $P = M \times G$ is the trivial bundle, then $Lie\ \cal G$
can be identified with $Map(M,\gothic g^*)$.

\vskip2ex
        Meanwhile, also around 1975, Fradkin and Vilkovisky \c {FV}
initiated a different approach to quantization in the
Hamiltonian setting.  They started with a phase space (=
symplectic manifold $W$, e.g. the cotangent bundle $T^*\cal A$)
and {\bf constraints}
$\phi_{\alpha}: W\rightarrow \Bbb R$ .  Via the symplectic structure, these
corresponded to Hamiltonian vector fields which were assumed
tangent to and foliating the constraint `surface', the zero
locus of the constraints.  They proceeded by adjoining
`ghosts', i.e. Grassmann algebra generators, to the Poisson
algebra of smooth functions on $W$ and defining an operator
$D$ on the extended algebra such that $D^2 = 0$.  The operator
$D$ contained a piece which was the Chevalley-Eilenberg differential
$d$ used in defining Lie algebra cohomology.  At least in nice
cases, the resulting cohomology in degree zero gave the algebra
of functions on the reduced phase space.

\vskip2ex
        By 1977, BFV (Batalin, Fradkin and Vilkovisky) saw their work
as a variant of BRST.  Since
then the term `BRST' has been applied to an ever increasing range of
situations - primarily quantum but also classical - in which
there is an operator $Q$ of square zero and hence `BRST
cohomology' $H =Ker Q/ Im Q$. I would like to restrict the term to
situations in which $Q$ contains a piece which is specifically of
Cartan-Chevalley-Eilenberg type.  This still includes one of the
most important variants in the infinite dimensional case - the
semi-infinite cohomology of Feigin \c F.
\vskip2ex
        Although many of the variants of BRST cohomology have
been developed for use in quantum field theory, they do have
classical analogs which are of considerable interest in their
own rite and help to reveal the relevance of cohomology, both
classical and quantum.  I will focus on one particular example,
the Batalin-Fradkin-Vilkovisky approach to first class
constrained Hamiltonian systems.  These systems are particularly
appropriate for this conference, their structure as mathematics
has become particularly clear and I can present them as more
than a spectator.  I will present only the classical aspects.
Since these can be expressed entirely in terms of differential
algebra, although of a new kind, algebraic techniques for
quantization apply quite well; see in particular work of
Huebschman \c {Hu} and also of Figueroa-O'Farrell and Kimura \c {FK}\c {Ki}
for the latest results and excellent expositions of the subject.
After presenting the classical aspects in some detail, I will sketch
several other related areas of what Witten has recently blessed
with the name `cohomological field theory' \c {Wi}.
\vskip2ex
        As mentioned, the Hamiltonian setting refers to
functions on a symplectic manifold, proto-typically a cotangent
bundle.  For present purposes,however, it is sufficient to consider a
Poisson algebra,
the formal algebraic object modeled on the algebra of smooth
functions on a symplectic manifold.  In light of the mixed
audience today, let me present a tri-lingual dictionary:
\vfill\eject

 $$\matrix
 \format\l&\qquad\c&\qquad\r\\
 \text{PHYSICS} &\text{GEOMETRY} &\text{ALGEBRA}\\
 \vspace{1\jot}\\
 \vspace{1\jot}\\
 \text{Hamiltonian system} &\text{Differential
 system}\\
 &\text{on symplectic }W \\
 &\text{ e.g. }T^\ast M\\
 \vspace{1\jot}\\
\vspace{1\jot}
 \text{Fields} &\text{Functions on }W &\text{Poisson
 algebra }P\\
 \text{with Poissen bracket}&\text{with Poisson
 bracket}\\
 \vspace{1\jot}\\
 \vspace{1\jot}
 \text{Constraints}\ \{ \varphi _\alpha\} &\phi :W\to
 \Bbb R^N &\text{Ideal }I\subset P\\
 \varphi_{\alpha}\in \Cal C^{\infty}(W)\\
 \vspace{1\jot}\\
 \vspace{1\jot}
 \text{Constraint surface} &V=\phi^{-1} (0)\subset W
 &P/I\\
 \vspace{1\jot}\\
 \vspace{1\jot}
 f\approx g\text{ (weakly equal)} &f\vert V\,=\,g\vert V
 &f\equiv g \mod I\\
 \vspace{1\jot}\\
 \vspace{1\jot}
 \text{Symmetries} &\text{Hamiltonian v.f.} &\text{ad
 action of } I\\
 \{ \varphi _\alpha,\ \} &X_{\varphi _\alpha}\\
 \vspace{1\jot}\\
 \vspace{1\jot}
 1^{st}\ \text{Class:}\\
 \{ \varphi _\alpha,\varphi _\beta\}=\Cal
 C^{\gamma}_{\alpha\beta} \varphi_\gamma &X_{\varphi_\alpha}\ \text{tangent
 to}\ V &I\ \text{closed under}\ \{ \ ,\ \}\\
 \text{structure {\bf functions}}
 &\text{and foliating} &\text{Lie algebra over} \Bbb R\\
 &&\text{{\bf not} over}\ P\\
 \vspace{1\jot}\\
 \vspace{1\jot}
 \text{true degrees of freedom} &\text{reduced phase
 space }V/\frak F  &( P/I)^{\text{I-invariant}}
 \endmatrix
$$
\vskip4ex
 \subheading{Definition}  A Poisson algebra $P$ over a
 field $k$ is a vector space $P$ over $k$ with two
 operations, denoted respectively by $f,g \rightarrow
 fg$ and $f,g \rightarrow \{f,g\}$, satisfying the
 following three conditions:
 \roster
 \item the product $fg$ makes P an associative algebra;
 \item the bracket $\{f,g\}$ makes P a Lie algebra;
 \item the two are related by a Leibnitz rule: $\{f,gh\} =
 \{f,g\}h + g\{f,h\}$\newline (otherwise said: $\{f,\ \}$ acts as
 a derivation of the associative algebra $P$.)
 \endroster

 It is common to assume that $P$ is in fact
 commutative, i.e. $fg =gf$, and we shall in fact do so,
 but the additional generality as stated is appropriate
 in light of both  quantization and current interest in
 non-commutative geometry.
\vskip2ex
A typical Hamiltonian
 system consists of differential equations of the form
 $\{f, H\} = df/dt$ where $H$ is a fixed function on the
 manifold $W$.  In some physical problems, solutions are
 sought which are constrained to lie on a sub-manifold
 $V \subset W$.  As in algebraic geometry, we can think
 of $V$ as the zero set of some functions $\phi_\alpha
 :W \rightarrow R$, called {\bf constraints}.  The
 algebra of functions
 $C^{\infty}$-in-the-sense-of-Whitney on $V$ can be
 identified with $C^{\infty} (W)/I$ where I is the ideal
 of functions which vanish on $V$. We restrict attention
 to situations in which the $\phi_{\alpha}$ generate
 $I$.  Example: Zero angular momentum.
 $$
 \align W &= \Bbb R^2 \times \Bbb R^2\\
  (P,Q)&=
 (q^1,q^2) \times (p_1,p_2)\\
 \phi &=
  P \otimes Q = p_1q^2 - p_2q^1 .\endalign
 $$
 \vskip2ex Now if $W$ is symplectic (or just given a
 Poisson bracket on $C^{\infty} (W))$, Dirac \c {D1} calls the
 constraints {\bf first class} if $I$ is closed under the
 Poisson bracket.  (If the $\Bbb R$-linear span $\Phi$ of
 the $\phi_{\alpha}$ is closed under bracket, physicists
 say the $\phi_{\alpha}$ {\bf close} on a Lie algebra;
 this is a very nice case, but the more general FIRST
 CLASS case in which the ideal $I$ is closed but $\Phi$
 is not is where homological techniques are really
 important.)  When the constraints are first class, we
 have that the Hamiltonian vector fields
 $X_{\phi_\alpha}$ determined by the constraints are
 tangent to $V$ (where $V$ is smooth) and give a
 foliation $\Cal F$ of $V$.  Similarly, $C^{\infty}(W)/I$
 is an $I$-module with respect to the bracket.  (In
 symplectic geometry, the corresponding variety is
 called {\bf coisotropic} \c {We}.)

\vskip2ex
An example of the special case that is particularly relevant to
this conference is that of an equivariant moment map \c {AGJ}.  Here we
are given a Lie group $G$ acting on $W$ by symplectic
diffeomorphisms (symplectomorphisms) and an equivariant {\bf moment} map
$$
       J = \Phi: W \rightarrow \gothic g*,
$$
equivariant with respect to the coadjoint action of $G$ on $\gothic
g^*$, the dual of $\gothic g$, the Lie algebra of $G$.  In the physics
literature, it is common to choose a basis $\lbrace
T^{\alpha}\rbrace$ and to write $\phi(w) = \phi_{\alpha}T^{\alpha}$.

\vskip2ex
In many cases of interest, $I$ does not arise as the Lie algebra
of some Lie group of transformations of $W$ or even $V$, but the
corresponding Hamiltonian vector fields $X_{\phi_\alpha}$ are
still referred to as (infinitesimal) symmetries.  In the nicest
case, e.g. when the foliation $\cal F$ is given by a principal
$G$-bundle structure on a smooth $V$, the algebra $C^{\infty}(V/\cal F)$
 can be identified with the $I$-invariant sub-algebra of
$C^{\infty}(W)/I$.  In great (if not complete) generality, this
$I$-invariant sub-algebra represents the true observables of the
constrained system.
Sniatycki and Weinstein \c {SW} have defined an algebraic
      reduction in the context of group actions and momentum maps which
      is guaranteed to produced a reduced Poisson algebra but not
      necessarily a reduced space of states.  The S-W (Sniatycki and
      Weinstein) reduced Poisson algebra is $ (C^{\infty} (W)/I)^G$   where
$V =  J^{-1}(0)$  for some equivariant moment map  $J : W \rightarrow \frak
g^*$.
 (If  $G$  is compact, $(C^{\infty} (W)/I)^G$   is
isomorphic to the Dirac reduction  $C^{\infty} (W)^G /I^G$ .)  With hindsight,
the generalization of S-W reduction to a general FIRST CLASS constraint ideal
$I$  is obvious.  The issue of its suitability is not one of geometry
necessarily, but rather one of physics.

\vskip2ex
Now - where are the ghosts?
Instead of considering just the ``observable'' functions, one
can consider the deRham complex of longitudinal or vertical
forms of the foliation $\cal F$, that is, the complex $\Omega (V,\cal F)$
consisting of forms on vertical vector fields. In
local coordinates $(x^1,...,x^{r+s})$ with $(x^1,...,x^r)$ being
coordinates on a leaf, a typical longitudinal form is
 $$ f_J (x)
dx^J \quad \text{where} \quad J = (j_1,...,j_q) \quad \text{with} \quad
1 \leq j_1 < ... j_q \leq r, \text{ the leaf dimension}.  $$

Another description of $\Omega(V,\cal F)$ is in terms of alternating functions
of vertical vector fields which are multi-linear with respect to
$C^{\infty}(V)$.  To become more fully algebraic, consider $P$, an arbitrary
Poisson algebra with an ideal $I$ which is closed under the Poisson bracket.
Reduction is then achieved by passing to the $I$-invariant
      subalgebra of  $P/I$.  Note that a class  $[g]$  is $I$-invariant if
      $\{I,g\} \subset I$, equivalently, if  $\{\phi,g\} \approx   0$  for all
constraints  $\phi \in I$. This subalgebra inherits a Poisson bracket even
though  $P/I$  does not:  For  $f,g \in P$  and
$\phi \in I, \{f + \phi,g\} = \{f,g\} + \{\phi,g\}$
      where  $\{\phi,g\}$   need not belong to  $I$, but will if the class of
$g$  is $I$-invariant.
\vskip1ex
           The fact that  $I$  is a sub-Lie algebra of  $P$  but is not a
  Lie algebra  over  $P$  (the bracket is {\bf R}-linear but not $P$-linear)
      is a significant subtlety.  The pair  $(P,I)$  is, however, a
      {\bf Rinehart algebra} \cite R  over {\bf R}  :
\vskip.5ex   $P$ is a commutative algebra over {\bf R},
 \vskip.5ex   $I$ is a Lie algebra over {\bf R} and a $P$-module,
\vskip.5ex $\{\phi,\quad \}$ gives a representation $\rho: I \rightarrow$ Der
$P$,
the Lie algebra of derivations of $P$,
such that $\{\phi,f\psi\} = (\rho (\phi)f)\psi + f\{\phi,\psi\}$
for $\phi,\psi \in I, f \in P$.
\vskip1ex\noindent
      Hence we can consider the {\bf Rinehart complex}  Alt${}_P (I,M)$  where
$M$ is a $P$-module with a representation $\pi$ as a Lie module over  $I$
      such that
$$
\pi(\phi)(fm) = f\pi (\phi)m + \pi(f\phi)m,\ \ f \in P, \phi
\in I, m \in M.
$$
The underlying vector space of Alt${}_P(I,M)$ consists of the alternating
$P$-multi-linear functions from $I$ to $M$. The differential  $d$  given by
Rinehart is an obvious generalization of that of Cartan-Chevalley-Eilenberg:
$$
(dh)(\phi_0,...,\phi_q) = \sum_{i < j} (-1)^{i+j}
h(\{\phi_i,\phi_j\},...,\hat \phi_i,...,\hat \phi_j,...) + \sum_i
(-1)^i \pi(\phi_i)h(...,\hat \phi_i,...).
$$
      (In case, $M = P = C^{\infty}(W)$  and  $I$ corresponds to  vector fields
on $W$, the Rinehart complex is the de Rham complex of $W$. )
\vskip1ex
When  $I$  is a subalgebra of FIRST CLASS constraints, $P$  is not a
$(P,I)$-module via the adjoint action:
$$
         \{\phi,fg\} = \{\phi,f\}g + f\{\phi,g\} \neq f\{\phi,g\} +
\{f\phi,g\},$$
but  $P/I$  is a $(P,I)$-module via the adjoint action since  $\{f\phi,g\}
\equiv f\{\phi,g\}$ mod $I$.  As remarked by Stephen Halperin, the Rinehart
complex  $Alt_P (I,P/I)$ is,
in this case,  the complex $\Omega^* (V,\Cal F)$  of longitudinal forms.
(See \cite {Hu} for further applications of Rinehart's complex to Poisson
algebras.)
\vskip2ex

In some special cases, what the physicists \c {BFV},  \c {He}, \c {BM}
 did was to construct a
homological ``model'' for $\Omega(V, \cal F)$ in roughly the sense
of rational homotopy theory \c {Su}.  That is to say, a differential
graded commutative algebra with a map to $\Omega(V, \cal F)$ giving
an isomorphism in cohomology.  The model was itself crucially a
Poisson algebra extension of the Poisson algebra $P =
C^{\infty}(W)$ and its differential contained a piece which
reinvented the Koszul complex for the ideal $I$.  The
differential also contained a piece which looked like the
Cartan-Chevalley-Eilenberg differential.

\vskip2ex
But still - where are the ghosts?  The ghosts are easiest to describe
and their meaning clearest in the case in which the ideal is
{\bf regular}. (At one time, regular ideals were known as Borel ideals.)
This is an algebraic condition, but implied by $I$ being the
defining ideal in $C^{\infty}(W)$ for $V = \phi^{-1}(0)$ when $0$ is
a regular value of $\phi: W \rightarrow {\Bbb R}^N$.

\vskip2ex
In order to construct a model of Alt${}_P (I,P/I)$, we reverse the
procedure of BFV and first provide a model for $P/I$ as a
$P$-module.  This model is a
differential graded commutative algebra $(P \otimes \Lambda\Psi,\delta )$
where $\Psi$ is a graded vector space (in fact, negatively
graded) and $\Lambda\Psi$   denotes the free graded
commutative algebra on the graded vector space $\Psi$.
(This grading is the opposite of the usual
convention in homological algebra, but is chosen to correspond
to the (anti-) ghost grading in the physics literature.) This
model is constructed as follows in terms of a set of constraints
(a more invariant description is given in HRCPA \c S):
Let $\lbrace\phi_{\alpha}\rbrace$ be a regular sequence of constraints
(physics: irreducible set), i.e., there are no relations of the form
$f^1\phi_1 + \dots + f^i\phi_i =0$ for non-zero $f^j$ in $P$.
Adjoin Grassmann variables, ghosts $\eta^{\alpha}$ and anti-ghosts
$\cal P_{\alpha}$ in 1-1 correspondence with the constraints.
That is, form the graded commutative algebra
$\Lambda\eta^{\alpha}\otimes P \otimes \Lambda\cal P_{\alpha}$.
Extend the Poisson bracket on P to this new algebra by decreeing
$$
\lbrace\eta^{\alpha}, \cal P_{\beta}\rbrace = \delta^{\alpha}_{\beta}
$$
and then apply the Leibnitz rule to determine the Poisson
bracket on general monomials in the ghosts and anti-ghosts.
Notice that we can interpret this bracket as follows: The span
of the $\phi_{\alpha}$ is isomorphic to $\Phi$ and the span of
the $\eta^{\alpha}$ is isomorphic to the dual, $\Phi^*$, so the
bracket formula above is the usual symplectic structure on $\Phi^*
\oplus \Phi$. The ghost degree is defined to be $1$ for $\eta^{\alpha}$
and $-1$ for $\cal P_{\alpha}$ and is extended to monomials `additively',
i.e. denoting ghost degree by $gh$, we have $gh(\omega_1\omega_2)
= gh \omega_1 + gh \omega_2$.
\vskip2ex
\subheading{Theorem} (Batalin-Fradkin-Vilkovisky): There exists $Q\in$ $\FBV$
of ghost degree $0$ such that $\lbrace Q,Q\rbrace = 0$.  The
operator $D= \lbrace Q,\quad\rbrace$ satisfies $D^2 = 0$ and in
ghost degree $0$, we have ${{\text{Ker} D}\over {\text{Im} D}} \approx
(P/I)^{I\text{-invariant}}$.

\vskip2ex\noindent
Many years later, after reinterpretation by Henneaux \c {He} and then
Browning and McMullan\c {BM}, I was able to reinterpret the existence
of $Q$ in terms of {\it Homological Perturbation Theory}.

\vskip2ex First consider just the anti-ghost complex, $P \otimes
\Lambda\cal P_{\alpha}$ with the derivation $\delta$ defined by
$\delta \cal P_{\alpha} = \phi_{\alpha}$.  Browning and McMullan
recognized this as the Koszul complex for the ideal $I$ in the
commutative algebra $P$ [K], [Bo].  The condition that the ideal
$I$ is regular is equivalent to the Koszul complex being a model
for $P/I$ or, in algebraists' terms, a resolution of $P/I$.  (For
more general ideals, this fails, i.e., $H^i (P \otimes
\Lambda\cal P_{\alpha},\delta) \neq 0$ for some $i \neq 0$.  The Tate
resolution \c {Ta} kills this homology by systematically enlarging
the set of anti-ghosts, cf. \c {FHST} and \c {S3}).

\vskip2ex
On the other hand, the ghost complex $\Lambda\eta^{\alpha}\otimes
P$ with the Chevalley-Eilenberg differential $d$ computes
$H_{Lie}(I,P)$ so that $H^0(I,P) = I$-invariants of $P$.
Similarly $\Lambda\eta^{\alpha}\otimes P/I, d$ computes the
$I$-invariants of $P/I$ in degree $0$.

\vskip2ex
The BFV differential $D$ looks like $d + \delta +$ terms of
higher order.  To construct it as an inner derivation
$D=\lbrace Q,\quad\rbrace$ with $Q = Q_0 + Q_1 +$ terms of
higher order, start with $Q_0 = \eta^{\alpha}\phi_{\alpha}$.
We then obtain the following formulas for the action of $Q_0$
on $P$ and on the ghost generators:
$$\align
\lbrace Q_0,f \rbrace &= \eta^{\alpha} \lbrace\phi_{\alpha},f
\rbrace = df\\
\lbrace Q_0, \etab \rbrace &= 0\\
\lbrace Q_0, \pb \rbrace &= \delta^\alpha_\beta\phia =
\delta\pb.
\endalign$$

\vskip2ex
Since $I$ is closed under Poisson bracket,
$$ \lbrace\phia,\phib\rbrace = C^{\gamma}_{\alpha
\beta}\phi_{\gamma}
$$
where the $C^{\gamma}_{\alpha \beta}$ may be {\it functions}.
Let $Q_1 = 1/2\etaa\etab C^{\gamma}_{\alpha
\beta}\phi_{\gamma}\phi_{\gamma}$ which then acts according to
the formulas:
$$\align
\lbrace Q_1, f\rbrace &= 1/2\eta^{\alpha}\etab \lbrace C^\gamma_{\alpha \beta}
,f \rbrace \pg\\
\lbrace Q_1, \etae \rbrace &= 1/2\etaa\etab\strfun = d\etae\\
\lbrace Q_1, \pe \rbrace &= \etaa C^\gamma_{\alpha\epsilon}\pg
\delta\pb.
\endalign$$
Thus $\lbrace Q_0 + Q_1, \quad\rbrace = d + \delta +$ stuff
where `stuff' stands for the terms above which do not appear
in $d + \delta$. Because of these extra terms, e.g. $\lbrace
Q_1, f\rbrace$, we have $\lbrace Q_0 + Q_1, Q_0 + Q_1 \rbrace
\neq 0$.  How can we add ``terms of higher order'' $Q_i$ so as
to achieve $D^2 = 0$?

\vskip2ex Here is the inductive step.  Note $Q_0$ has one ghost
and no anti-ghosts while $Q_1$ has two ghosts and one anti-ghost.
Assume we have defined $Q_i$ with $i+1$ ghosts and $i$ anti-ghosts
and that $$R_n:= \Sigma^n_0 Q_i $$ is such that $\lbrace R_n,
R_n\rbrace$ is a sum of terms, each of which has at least $n+1$
anti-ghosts ($\cal P's$) and that $\delta\lbrace R_n, R_n\rbrace$
has terms with at least one more.  Because $\delta$ is acyclic,
there is a suitable $Q_{n+1}$.  In fact, there is a {\it
contracting homotopy} for $\delta$, i.e. a linear map $h: P\otimes
\Lambda{\cal P} \rightarrow P\otimes \Lambda{\cal P}$ which raises
the number of anti-ghosts by $1$ such that $\delta h + h \delta =
Id - \bar\pi$ where $\bar\pi : P \otimes \Lambda\Psi \rightarrow P
\rightarrow P/I \hookrightarrow P \otimes \Lambda\Psi$ is given by $\pi$
composed with an {\bf R}-linear splitting $P/I\hookrightarrow P$.
Having made one such choice, we can then systematically choose
$$Q_{n+1} = -1/2 h\{R_n,R_n\}.$$
\vskip2ex
That we have constructed a model for (Alt${}_P (I,P/I),d)$
follows if we can show\newline $\pi:((\Lambda\Psi)^* \otimes P \otimes
\Lambda\Psi,\partial)\rightarrow \text{Alt}_P (I,P/I,d)$ induces a homology
isomorphism.  In the regular case, this follows by the usual techniques of
comparison in homological perturbation theory, namely comparison
of spectral sequences, but with some subtlety.

\vskip1ex
When the ideal $I$ is not regular or the chosen set of
constraints is reducible even though the ideal is regular, the
Koszul complex is not a resolution of $P/I$. It can however be
extended to the (Koszul)-Tate resolution \cite {Ta}, by adjoining
alternately new even and odd variables, called in physics
``anti-ghosts of anti-ghosts'' etc.  The Tate resolution again
admits a contracting homotopy, so the construction of $Q$
 proceeds as above.  If the set of constraints is reducible but
the ideal is regular (for example if the corresponding vector
fields define the action of a Lie group G but the orbits are in fact
homogeneous spaces $G/H$), the result is again a model
 for the deRham complex of forms along the leaves.  In
 essence, the extra constraints have led to a model
 containing a factor which is acyclic and so makes no
 contribution to the cohomology.  When the ideal is NOT
 regular, the cohomology in degree zero is still
 isomorphic to the $I$-invariants of $P/I$, but the
 interpretation of the other groups is less clear.  Even
though the ideal is not regular, the zero locus $V$ of the
constraints may still be a sub-manifold of $W$ and the
corresponding Hamiltonian vector fields may give a true
foliation (without singularities).  That is the case considered
in \c {FHST} where we show we again have a model
for the deRham complex of forms along the leaves.
Should the BFV complex have cohomology different from that of
the deRham complex of forms along the leaves, it is an interesting
question as to which cohomology is physically significant..
\vskip3ex
 \subheading{Ghosts in the Lagrangian setting} Problems
 in classical field theory are, if anything, more
 familiar in the Lagrangian setting.  We start with an
 ``action'' $S_0 = S_0(\phi)$ where $\phi$ denotes one
 or several `fields'. Nowadays the word seems to
 indicate a function  or section of some bundle $p:E\rightarrow M$.
 We seek solutions of a variational equation or system
 of equations
 $$
\delta S_0 := \frac{\delta}{\delta \phi}S_0 = 0.
 $$
 The point of view relevant to cohomological (ghost)
 techniques considers $\Sigma$, the space of all
 solutions as a subspace of the space $\Cal S = \ \text{Sections }E$
 of all fields.  The following discussion of these
 techniques is essentially just an introduction to Marc
 Henneaux's talk, which will provide a more thorough
 treatment.

 \vskip1ex
 As in the Hamiltonian setting, we begin with
 an algebra of functions, e.g. $C^{\infty}\Cal S$,
 although this time just a commutative algebra, not a
 Poisson algebra.  We ignore all difficulties associated
 with the infinite dimensional nature of $\Cal S$ and
 proceed with an algebraic formalism.  We approach the
 subspace $\Sigma$ via a Koszul complex.

 \vskip2ex Let $\phi^i$ be a (minimal) set of solutions
 of the variational equation or rather let $\phi^i$ be
 functions on $\Cal S$ corresponding to a minimal set of
 solutions generating all solutions.  Introduce a new
 set of Grassmann variables $\phi^*_i$ of ghost degree
 $-1$, called ``anti-fields'', to generate the Koszul
 complex:
 $$
 C^\infty \Cal S \otimes \Lambda\phi_i^\ast.
 $$
 Define a Poisson ``anti-bracket'', denoted $(\ ,\ )$,
 by declaring $(\phi^i,\phi^*_j) =
 \delta^i_j$ and extending according to a (slightly
 mis)graded Leibnitz rule.  Notice that the above
 formula holds in ghost degree $0$ but is applied to
 terms of ghost degree $-1$ and $0$ respectively; this
Interrupt
 (equivalent ways) of describing the Leibnitz rule:
 \roster
 \item  the anti-bracket itself is an operation of ghost degree
 1, i.e. (denoting total ghost degree by ``$gh$'')
 $gh(A,B) = gh A + gh B + 1$, so the Leibnitz rule is:
 $$
 (A,BC) = (-1)^c(A,B)C + (-1)^{b(a+1)}B(A,C);
 $$
 \item the anti-bracket is a graded Poisson bracket with
 respect to {\it degree} where degree $A=gh A+1$.\endroster
 \vskip2ex
 The variational system or, more importantly,
 the space of solutions $\Sigma$ may have Noether
 symmetries, i.e. may support the vector field action of
 a Lie algebra $\goth g$. Choose a basis $C^*_{\alpha}$
 for $\goth g$ declared to have ghost degree -2 and a
 dual basis $C^{\alpha}$ declared to have ghost degree 1
 for the dual of $\goth g^*$.  Define the anti-bracket
 of these new generators to be
 $$
 (C^{\alpha}, C^*_{\beta}) = \delta^{\alpha}_{\beta}
 $$
 and extend by the above Leibnitz rule. This notation
 is a mild modification of that in the physics
 literature: $C^*_{\alpha}$ is more traditionally denote
 $\phi^*_{\alpha}$.
 \vskip2ex What's going on here!  As Henneaux will
 explain more fully, the anti-fields $\phi^*_{\alpha}$
 can be interpreted as vector fields on $\Sigma$ and
 $(\quad ,\quad )$ as the Schouten bracket.  Here is a
 very preliminary attempt to interpret the curious
 regrading by degree.
 \vskip2ex Following Henneaux and others, interpret
 $\Cal S$ as a space of {\it histories}, which I take to
 mean sections of the bundle $E\rightarrow M$ which is
 of the form $D\times I\rightarrow N\times I$.  Thus
 sections can be interpreted as 1-parameter families of
 sections of $D\rightarrow N$ and $\Cal S$ as
 $$
 \text{Map}\ (I,
 \text{Sections }(D \rightarrow N)).
 $$
 There are homological
 models for such path spaces in the mathematical
 literature; hopefully one of them can be related
 straightforwardly to this anti-bracket model.
 \vskip2ex The major result of Batalin and Vilkovisky in
 this Lagrangian setting is that the appropriate action
 $S$ to be quantized is of the form $S = S_0 +$ \lq\lq ghost
 terms" where the ghost terms each have ghost degree $0$
 as does the $S_0$ with which we began the discussion.
 It should come as no surprise that the existence of the
 ghost terms follows from the acyclicity of the Koszul
 complex (assuming appropriate regularity conditions) or
 of the Koszul-Tate resolution.
 \subheading{Cohomological field theory}
 \vskip2ex Witten has recently \c {Wi} introduced the term
 ``cohomological field theory'' for yet another
 situation in which techniques of homological algebra
 play a significant role.  I would suggest that the term
 be applied to the Hamiltonian and Lagrangian methods
 above as well.
 \vskip2ex Witten's paper is concerned in passing with
 ``equivariant cohomology'', apparently inspired by
 discussions with Scott Axelrod. My remarks will provide
 an introduction to a small part of Axelrod's talk later
 this week.
 \vskip2ex Equivariant cohomology refers to cohomology
 in the setting of transformation groups, i.e. a
 topolgical group $G$ acting on a space $X$ without
 assumptions on the action $G\times X \rightarrow X$
 other than its continuity or smoothness (and the
 algebraic conditions that the unit of $G$ act as the
 identity on $X$ and that $g(hx) = (gh)x$ for $g,h\in G,
 x\in X$).  Without further restriction on the action,
 the quotient space (= orbit space) $X/G$ may fail to be
 a manifold or even Hausdorff.  The projection
 $X\rightarrow X/G$ need be nothing like a principal
 bundle.  Thus the cohomology of $X/G$ may be very
 difficult to relate to that of $X$ and $G$.
\vskip1ex
 Borel in his celebrated Transformation Group Seminar
 constructed a principal $G$-bundle $\tilde X\to X_G$ where $\tilde X$ has the
same homotopy type as $X$.  In case the action of $G$ on $X$ does give a
principal $G$-bundle $X\to X/G$, then $X_G$ has the homotopy type $X/G$ and
hence the same cohomology. Otherwise, $X_G$ is a space with the cohomology that
the orbit space {\it should} have. This cohomology is known as the {\bf
equivariant cohomology} of $X$, denoted $H_G(X):=H(X_G)$.

The construction of $X_G$ makes use of the {\it universal} principal $G$-bundle
$EG\to BG$. The space $\tilde X$ is in fact just $EG\times X$ with the diagonal
action of $G$ and $X_G$ is the orbit space $EG\times_GX$.

In the smooth setting, $G$ a Lie group acting smoothly on a manifold $M$, there
is a model for the equivariant cohomology which uses the Weil algebra, a model
for the differential forms on EG.  The $G$-action on $M$ is
reflected in two families of operators $\iota_X$ and $\theta_X$ on
differential forms on $M$ for $X\in \frak g$ (and similarly for the
$G$-action on $EG$).
For $X\in\frak g$, $\iota_X$ is contraction with $X$ and $\theta_X$ is the
infinitesimal action of the vector field corresponding to $X$.
These operators obey the usual rules:
$$\align
\iota_{[X,Y]} &= \theta_X\iota_Y - \iota_X \theta_Y\\
\theta_X &= d\iota_X + \iota_X d.
\endalign$$
Now construct the  {\bf Weil algebra}
$= W(\frak g) := \Lambda(\frak g^\ast)\otimes S(s\frak g^\ast)$,
the free graded commutative algebra
generated by a copy of the dual of $\frak g^\ast$ in degree 1 and
another copy $s\frak g^\ast$ of $\frak g^\ast$ in degree 2.
 (If we chose a basis $C^\alpha$ for
$\frak g^\ast$ in degree , the Weil algebra would be described as generated by
the ghosts  $C^\alpha$ with ghost degree 1 and by even generators $sC^\alpha$
with ghost
degree 2.)  This algebra $W(\frak g)$ is given a total differential D of degree
1 which is the sum of two differentials $\delta$ and $s$ where $\delta$ is the
Chevalley-Eilenberg differential for $\frak g$ with coefficients in $S(s\frak
g^\ast)$ under the coadjoint action and $s$ is the differential determined on
the odd generating copy of $\frak g^\ast$ as an isomorphism with the even copy
(in coordinates, $s:C^\alpha \to sC^\alpha)$.  Alternatively, $s$ can be
regarded as isomorphic to the graded analog of the Koszul differential for
the maximal ideal of $S(s\frak g^\ast)$.
Since this algebra is acyclic with respect to $s$,
it is also acyclic with respect to $D=s+\delta$.  As such, it is a model for
(has the same cohomology as) $EG$.  The principal $G$-action on $EG$ is
reflected in $W(\frak g)$ by defining the operators $\iota_X$ and $\theta_X$
for $X\in \frak g$ as follows:
$$\align
\theta_X(h) &= -h([X,\quad])\quad \text {for} \quad h\in \frak g^\ast\\
\theta_X(sh) &= s\quad coad(X)h
\endalign$$
where $coad$ denotes the coadjoint action of $\frak g$ on $\frak g^\ast$ and
$$
\iota_X h = h(X) \quad \text {for} \quad h\in \frak g^\ast
$$
while $\iota_X$ is $0$ on $S(s\frak g^\ast)$.
These operators combine to give operators $\iota_X$ and $\theta_X$
on $W(\frak g)\otimes \Omega^\ast(M)$.  The cohomology of $M_G$ is
then given by the subcomplex of
\lq\lq basic" forms:  Ker $\iota_X \cap \text{ Ker }\theta_X$.
\vskip2ex
Since $\iota_X$ is 0 on $S(s\frak g^\ast)$ and non-degenerate on
$ \Lambda(\frak g^\ast)$, the ``basic'' forms lie, in fact, in
$S(s\frak g^\ast)\otimes \Omega^\ast(M)$.  The algebra is reflecting the
homotopy fibration $X\rightarrow X_G \rightarrow BG$ where $BG$ is
the classifying space of the group $G$.  Thus the `bosonic' ghosts
$sC_{\alpha}$ come `from below'.
Axelrod will give some indication in his talk of how this construction is
relevant to classical field theory (see also \c A).
\vskip3ex
\centerline{\bf References}
\vskip3ex

	[A]    	S. E. Axelrod, {\sl Geometric Quantization of Chern-Simons Gauge
                 Theory}, Dissertation, 1991.

\vskip.5ex      S. E. Axelrod and I. M. Singer, Chern-Simons perturbation
theory,
                {\sl Proceedings of the XXth Conference on Differential
		Geometric Methods in Physics}, Baruch College/CUNY,
		New York, NY, 1991.

\vskip.5ex
      [AGJ]      J.M. Arms, M. J. Gotay and G. Jennings, (Geometric and
                 Algebraic) Reduction for Singular Momentum Maps, preprint.
\vskip.5ex
\flushpar {\it [BFV] see also [FF] and [FV]}
\vskip.5ex
      [BF]       I.A. Batalin and E.S. Fradkin, A generalized canonical
                 formalism and quantization of reducible gauge theories,
                 Phys. Lett. 122B (1983), 157-164.
\vskip.5ex
      [BV1]      I.A. Batalin and G.S. Vilkovisky, Existence theorem for
                 gauge algebra, J. Math. Phys. 26 (1985), 172-184.
\vskip.5ex
      [BV2]      I.A. Batalin and G.S. Vilkovisky, Quantization of gauge
                 theories with linearly dependent generators, Phys. Rev. D
                 28 (1983), 2567-2582.
\vskip.5ex
      [BV3]      I.A. Batalin and G.S. Vilkovisky, Relativistic S-matrix
                 of dynamical systems with boson and fermion constraints,
                 Phys. Lett. 69B (1977), 309-312.
\vskip.5ex
      [BRS]	 C. Becchi, A. Rouet and R. Stora, Renormalization of the
                 abelian Higgs-Kibble model, Commun. Math. Phys. 42 (1975),
127.

\vskip.5ex       C. Becchi, A. Rouet and R. Stora, Renormalization of gauge
                 theories, Ann. Phys. 98 (1976), 287.

\vskip.5ex
      [BCR]	 L. Bonora and P. Cotta-Ramusino, Some remarks on BRS
		transformations, anomalies and the cohomology of the Lie
		algebra of the group of gauge transformations, Comm. in
		Math. Physics 87 (1983), 589-605.

\vskip.5ex
      [Bo]       A. Borel,  Sur la cohomologie des espaces fibr\'es principaux
et des espaces homogenes de groupes de Lie compacts, Annals of Math. 57 (1953),
115-207.
\vskip.5ex
      [BM]       A.D. Browning and D. McMullen, The Batalin, Fradkin,
                 Vilkovisky formalism for higher order theories, J. Math.
                 Phys. 28 (1987), 438-444.
\vskip.5ex
      [CE]       C. Chevalley and S. Eilenberg, Cohomology theory of Lie
                 groups and Lie algebras, Trans. Amer. Math. Soc. 63
                 (1948), 85-124.
\vskip.5ex

      [D1]        P.A.M. Dirac, {\sl Lectures on Quantum Mechanics}, Belfer
                 Graduate School Monograph Series 2 (1964).

\vskip.5ex
     [D2]       P.A.M. Dirac, Quantised singularities in the electromagnetic
		field, Proc. Royal Soc. A 133 (1931), 60-72.

\vskip.5ex
      [F]	 B. Feigin,
 The semi-infinite homology of Kac-Moody and Virasoro Lie algebras,
 Russian Math Surveys,
 39 (1984), 155-156; Russian original, Usp. Mat. Nauk 39 (1984), 195-196.

\vskip.5ex
 [FK]   J. M. Figueroa-O'Farrill and T. Kimura,
 Geometric BRST Prequantization I: Prequantization,
 Comm. in Math. Phys. 136 (1991), 209-229.

\vskip.5ex
 J. M. Figueroa-O'Farrill and T. Kimura,
 Homological Approach to Symplectic Reduction,
 University of Texas/Leuven preprint, 1991.

\vskip.5ex
      [FHST]     J. Fisch, M. Henneaux, J. Stasheff and C. Teitelboim,
                 Existence, uniqueness and cohomology of the classical
                 BRST change with ghosts of ghosts, Comm. Math. Phys. 120
		 (1989), 379-407.

\vskip.5ex
      [FF]       E.S. Fradkin and T.E. Fradkina, Quantization of relativistic
system with boson and fermion first- and second-constraints, Phys. Lett.
72B (1978), 343-348.

\vskip.5ex
      [FV]       E.S. Fradkin and G.S. Vilkovisky, Quantization of
                 relativistic systems with constraints, Phys. Lett. 55B
                 (1975), 224-226.

\vskip.5ex
      [G1]       V.K.A.M. Gugenheim, On a perturbation theory for the
                 homology of a loop space, J. Pure and Appl. Alg. 25
                 (1982), 197-205.
\vskip.5ex
      [GM]       V.K.A.M. Gugenheim and J.P. May, On the theory and
                 application of torsion products, Memoirs Amer. Math. Soc.
                 142 (1974).
\vskip.5ex
      [GS]      V.K.A.M. Gugenheim and J.D. Stasheff, On perturbations
                 and $A_{\infty}$-structures, Bull. Soc. Math. Belg. 38 (1986),
                 237-245.

\vskip.5ex
      [HH]	 J. L. Heitsch aand S. E. Hurder, Geometry of foliations,
		 J. Diff. Geom. 20 (1984), 291-309.
\vskip.5ex
      [He]        M. Henneaux, Hamiltonian form of the path integral for
                 theories with a gauge freedom, Phys. Rev. 126 (1985),
                 1-66.

\vskip.5ex
 [Hu] J. Huebschmann,
 Extensions of Lie Algebras and Poisson Cohomology,
 Heidelberg preprint, 1989.

\vskip.5ex
  J. Huebschmann, Graded Lie-Rinehart algebras, graded Poisson algebras, and
 BRST-quantization I: The Finitely Generated Case, Heidelberg preprint, 1991.

\vskip.5ex
 [Ki]
 T. Kimura,
 Generalized Classical BRST and Reduction of Poisson Manifolds,
(To be submitted to Comm. Math. Phys.).

\vskip.5ex
 T. Kimura,
 Generalized Geometric BRST Prequantization,
 (Work in Progress).

\vskip.5ex
      [Ko]        J.-L. Koszul, Sur un type d'algebres differentielles
                  en rapport avec la transgression, {\it Colloque de
Topologie},
                  Bruxelles (1950), CBRM, Liege.
\vskip.5ex
      [MW]       J.E. Marsden and A. Weinstein, Reduction of symplectic
                 manifolds with symmetry, Rep. Math. Phys. 5 (1974),
                 121-130.

\vskip.5ex
      [R]        G. Rinehart, Differential forms for general commutative
                 algebras, Trans. Amer. Math. Soc. 108 (1963), 195-222.
\vskip.5ex
      [SW]       J. Sniatycki and A. Weinstein, Reduction and
                 quantization for singular momentum mappings, Lett. Math.
                 Phys. 7 (1983), 155-161.
\vskip.5ex
      [S1]       J.D. Stasheff, Constrained Hamiltonians: A homological
                 approach, Proc. Winter School on Geometry and Physics,
                 Suppl. Rendiconti del Circ. Mat. di Palermo, II, 16
                 (1987), 239-252.
\vskip.5ex
      [S2]       J.D. Stasheff, Constrained Poisson algebras and strong
                 homotopy representations, Bull. Amer. Math. Soc. (1988),
                 287-290.

\vskip.5ex
           [S3]  J.D. Stasheff,
 Homological reduction of constrained Poisson algebras,
 to appear in
J.~Diff. Geom.

\vskip.5ex
      [Su]       D. Sullivan, Infinitesimal computations in topology,
  Math. IHES 47 (1977), 269-331.

\vskip.5ex
      [Ta]        J. Tate, Homology of Noetherian rings and local rings,
                 Ill. J. Math. 1 (1957), 14-27.
\vskip.5ex
      [Ty]	 I.V. Tyutin, Gauge invariance in field theory and statistical
		physics in operator formulation (in Russian), Lebedev Physics
		Inst. preprint No. 39 (1975).

\vskip.5ex
      [We]	 A. Weinstein, Coisotropic calculus and Poisson groupoids,
		 J. Math. Soc. Japan 40 (1988), 705-727.
\vskip.5ex
     [Wi]       E. Witten, Cohomological field theories,
                 Internat'l. J. Modern Physics A 6 (1991), 2775-2792.

\vskip.5ex
	[Y]	  C. N. Yang, {\sl Selected Papers, 1945-1980 (With Commentary)},
                  W. H. Freeman and Company, New York, pp. 73-74.
\end